\documentclass{article}

\usepackage{amsmath,amssymb,mathrsfs}
\usepackage{authblk}
\usepackage{hyperref}
\hypersetup{backref,bookmarksopen=true,colorlinks=true,pdfstartview=FitH,citecolor=blue,runcolor=blue,anchorcolor=blue,linkcolor=blue,
pdftitle = {UandR},pdfcreator = {\LaTeX\ with package \flqq hyperref\frqq},pdfproducer = {pdfLaTeX-\the\pdftexversion\pdftexrevision}}

\let\oldabstract\abstract
\let\oldendabstract\endabstract
\makeatletter
\renewenvironment{abstract}
{%
               {\list{}{\addtolength{\leftmargin}{1em} 
                        \listparindent 1.5em%
                        \itemindent    \listparindent%
                        \rightmargin   \leftmargin%
                        \parsep        \z@ \@plus\p@}%
                \item\relax}%
               {\endlist}%
\oldabstract}
{\oldendabstract}
\makeatother

\numberwithin{equation}{section}

\newcommand{\lag}{\mathscr{L}}

\begin{document}

\title{Unitarity and Renormalizability - Together at last}

\author{G B Tupper\footnote{Email: gary.tupper@uct.ac.za} \footnote{This paper originally appeared in the Proceedings of the 27th Annual Seminar on Theoretical Physics, Johannesburg,  29 June – 3 July 1992, ISBN 0 7972 0436 9. It is being released on the web to make the background field methodology for spontaneously broken theories available to a wider audience.}}

\affil{Department of Physics, University of Cape Town \\ Associate member, NITheP}

\date{\today}

\maketitle

\begin{abstract}
By construction, gauge theories require gauge fixing. In conventional approaches to spontaneously broken gauge theories, the choice of the Unitary ('t Hooft) gauge involves the sacrifice of manifest renormalizability (unitarity). It is shown that with a suitable modification of the background field gauge condition, the background field formalism allows manifest unitarity and renormalizability in a single framework.
\end{abstract}

\section{Introduction}
The development of spontaneously broken gauge theories represents a watershed in the progress of theoretical physics. Such theories form not only the basis for the enormously successful Standard Model but also for the various attempts at a grand unified theory of strong, weak and electromagnetic interactions.

While the basic principles were laid down by Higgs \cite{Higgs} and Kibble \cite{Kibble} in the early 1960's, the then available technology for quantizing the broken theory was based upon the 	`unitary (U-) gauge'. As it will figure prominently in our considerations let us briefly recall the U-gauge formalism.

I assume here a simple gauge group $\underline{G}$, Lie algebra $\mathcal{G}$ with representatives of generators $t_a$ and structure constants $C_{abc}:\left[ {{t_a},{t_b}} \right] = i{C_{abc}}{t_c}$. I also denote by $T_c$ the adjoint representation $(T_c)^{ab}=-iC_{abc}$. Neglecting matter fields the general model is 
\begin{align}
\lag\left( {A,\phi } \right) = {\lag_{YM}} + {\lag_H} =  - \frac{1}{4}F_{\mu \nu }^a\left( A \right){F^{a\mu \nu }}\left( A \right) + \frac{1}{2}\phi _{;\mu }^T{\phi ^{;\mu }} - {V_c}\left( {{\phi ^T}\phi } \right)
\label{eqn8}
\end{align}
where 
\begin{align}
F_{\mu \nu }^a\left( A \right) &= {\partial _\mu }A_\nu ^a - {\partial _\nu }A_\mu ^a - g{C_{abc}}A_\mu ^bA_\nu ^c\\
{\phi _{;\mu }} &= {D_\mu }\left( A \right)\phi  = {\partial _\mu }\phi  + ig{t_a}A_\mu ^a\phi \\
{V_c}\left( {{\phi ^T}\phi } \right) &= \frac{1}{2}{\mu ^2}{\phi ^T}\phi  + \frac{1}{4}\lambda {\left( {{\phi ^T}\phi } \right)^2} \label{eqn9}
\end{align}
Note that the Higgs field $\phi$ is taken to form a real representation - e.g. for $G=U(1), \phi^T=(\phi_1,\phi_2)$ and $t=-\sigma^2$. $\lag$ is invariant under the infinitesimal transformations 
\begin{align}
\delta \phi  =  - i{\theta _a}{t_a}\phi  \qquad \delta A_\mu ^a = \frac{1}{g}D_\mu ^{ab}\left( A \right){\theta _b}
\end{align}
with
\begin{align}
D_\mu ^{ab}\left( A \right) = {\delta ^{ab}}{\partial _\mu } + ig{\left( {{T_c}} \right)^{ab}}A_\mu ^c
\end{align}
the gauge covariant derivative with respect to $A$ in the adjoint representation. 

When $\mu^2 <0$ the invariance of $\lag$ is spontaneously broken through a non-trivial vacuum expectation value of the Higgs field
\begin{align}
{\left. {\frac{{\partial {V_c}}}{{\partial {\phi _{i}}}}} \right|_{\phi  = v}} = 0 \qquad {\left. {\frac{{{\partial ^2}{V_c}}}{{\partial {\phi _{i}}\partial {\phi _{i}}}}} \right|_{\phi  = v}} = {\left( {{M_{cs}}^2} \right)_{ij}} \ge 0
\end{align}
or in the case at hand ${\phi ^T}\phi  = {v^2} =  - \frac{{{\mu ^2}}}{\lambda }$. In general $G$ may have a subgroup $G'$ such as to 
leave the vacuum invariant: $t_av=0$ if ${t_a} \in \mathcal{G'}$. The U-gauge transformation is then 
\begin{subequations}
\begin{align}
&\phi  \to U\left( \xi  \right)\left( {v + \eta } \right) = \exp \left( { - i{\xi ^a}{t_a}/v} \right)\left( {v + \eta } \right), \quad {t_a} \in \mathcal{G}/\mathcal{G'}\\
&\left( {{t_a}A_\mu ^a} \right) \to U\left( \xi  \right){t_a}A_\mu ^a{U^\dag }\left( \xi  \right) - \frac{i}{g}U\left( \xi  \right){\partial _\mu }{U^\dag }\left( \xi  \right)
\end{align}
\end{subequations}
yielding 
\begin{align}
\lag \to  - \frac{1}{4}F_{\mu \nu }^a\left( A \right){F^{a\mu \nu }}\left( A \right) + \frac{1}{2}A_\mu ^a{\left( {M_{cV}^2} \right)^{ab}}{A^{b\nu }} + \frac{1}{2}\left[ {\eta _{,\mu }^T{\eta ^{,\mu }} + {\eta ^T}M_{cs}^2\eta } \right] +  \ldots
\end{align}
Here ${\eta _{,\mu }} \equiv {\partial _\mu }\eta $, the ellipsis stands for $\eta - A$ and $\eta$-self interaction terms \cite{c3} while 
\begin{align}
{\left( {M_{cV}^2} \right)^{ab}} = {g^2}v{t_a}{t_b}v
\end{align}
is the vector mass matrix. The would-be Goldstone bosons -- the angles $\xi$ parametrizing the coset $G/G'$ --  have disappeared from $\lag$, being replaced by the longitudinal degree of freedom of the massive vector propagator
\begin{align}
i{D_{\mu \nu }}\left( {k,m} \right) = \frac{i}{{{k^2} - {m^2}}}\left[ { - {g_{\mu \nu }} + \frac{{{k_\mu }{k_\nu }}}{{{m^2}}}} \right]
\label{eqn111}
\end{align}

As the name implies the U-gauge makes unitarity manifest in that only the physical degrees of freedom appear. At the same time for euclidean loop momentum $\ell_E$, $D\left( {{\ell _E} \to \infty ,m} \right) \to \mathcal{O}\left( 1 \right)$ so the transformed theory is \textbf{not} renormalizable by power counting. This does not mean that unitarity is lost any more than the symmetry is broken; rather both are well hidden. Following Weinberg one says that the U-gauge is `crypto-renormalizable'.

A modified quantization procedure enabling the renormalization of broken gauge theories was first introduced by 't Hooft \cite{tHooft1} and elaborated by Lee and Zinn-Justin \cite{blee}. In these renormalizable or $R_\xi$ gauges \cite{fujikawa} one begins by shifting the Higgs field $\phi \to v+\phi'$.  The kinetic piece $1/2\phi _{;\mu }^T{\phi ^{;\mu }}$ then gives rise to a term 
\begin{align*}
\frac{1}{2}igA_\mu ^a{\partial ^\mu }\left( {\phi {'^T}{t_a}v - {v^T}{t_a}\phi '} \right)
\end{align*}
which mixes the gauge and would be Goldstone fields. In order to remove this mixing one chooses a gauge fixing condition ${f^a}\left( {A,\phi } \right) = 0$ with 
\begin{align}
{f^a}\left( {A,\phi } \right) = {\partial ^\mu }A_\mu ^a - i\xi \phi {'^T}{t_a}v
\end{align}
so it is cancelled by
\begin{align}
{\lag_{GF}} =  - \frac{1}{{2\xi }}\left( {{\partial ^\mu }A_\mu ^a - i\xi g\phi {'^T}{t_a}v} \right)\left( {{\partial ^\nu }A_\nu ^a - i\xi g v{^T}{t_a}\phi'} \right)
\end{align}
Now both the scalar fields and Faddeev-Popov ghosts acquire gauge parameter dependent mass matrices
\begin{subequations}
\begin{align}
{\left( {\bar M_{cs}^2} \right)_{ij}} &= {\left( {{M_{cs}}} \right)_{ij}} + \xi g_i^2\left( {{t_a}v} \right)\left( {{v^T}{t_a}} \right)\\
{\lag_{FPG}} &= C_a^\dag {\partial ^\mu }D_\mu ^{ab}\left( A \right){C_b} + C_a^\dag \xi {\left( {M_{cV}^2} \right)^{ab}}{C_b}
\end{align}
\end{subequations}
but the massive gauge field propagator 
\begin{align}
i{D_{\mu \nu }}\left( {k,m} \right) = \frac{i}{{{k^2} - {m^2}}}\left[ { - {g_{\mu \nu }} + \left( {1 - \xi } \right)\frac{{{k_\mu }{k_\nu }}}{{{k^2} - \xi {m^2}}}} \right]
\label{eqn115}
\end{align}
is well behaved, $D\left( {{\ell _E} \to \infty ,m} \right) \to O\left( {\ell_E^{ - 2}} \right)$. Hence the $R_\xi$ gauge is manifestly renormalizable by power counting. 

Here the price one pays is that the would-be Goldstone bosons are still around as spurious degrees of freedom. What must be shown through the use of Generalized Ward Identities (GWI's) is that they as well as the other gauge dependent parts cancel in the S-matrix. Thus unitarity is \textbf{not} manifest in the $R_\xi$ gauges; again it is not lost but only hidden. One might sat the $R_\xi$ gauge is `crypto-unitary'. 

While the conventional renormalizable gauges are sid pro quo, it is natural to inquire whether all this cryptology is necessary, or whether it is not possible to instead achieve manifest renormalizability and unitarity in a single framework. In this paper I will show that the answer is indeed yes. The resolution of the schism lies in a modification of the Background Field Gauge (BFG) formalism first developed by De Witt \cite{bdw} for quantization of unbroken Yang-Mills theories and general relativity through one-loop. Subsequently the extension beyond the one-loop level to all orders was made by 't Hooft \cite{tHooft2} and Abbott \cite{labb1,labb2}.

As the background field method remains relatively obscure in Section 2 I will briefly review the formalism for scalar fields and global symmetries following the lucid presentation of Abbott \cite{labb2}; there in I also illustrate its simplification of the proof of renormalizability. Then in Section 3 I review the BFG for unbroken gauge theories. In Section 4 I return to the problem of spontaneously broken gauge theories and show how a modified BFG leads to manifest unitarity and renormalizability. Finally Sections 5 presents some calculations.

\section{The Background Field Formalism}
Given a scalar $\phi_i$ field system and lagrangian $\lag(\phi)$ such as $\lag_H$ above the familiar quantization procedure is to introduce external sources $J_i$ and the functionals
\begin{align}
W\left[ J \right] &= {\mathcal{N}^{ - 1}}\int {\left[ {d\phi } \right]\exp \left\{ {i\int {{d^4}x\left[ {\lag\left( \phi  \right) + {J_i}{\phi _i}} \right]} } \right\}}
\end{align}
\begin{subequations}
\begin{align}
Z\left[ J \right] &=  - i\ln W\left[ J \right] = \Gamma \left[ \Phi  \right] + \int {{d^4}x{J_i}{\Phi _i}} \\
{\Phi _i} &= \frac{{\delta Z}}{{\delta {J_i}}} , \qquad {J_i} =  - \frac{{\delta \Gamma }}{{\delta {\Phi _i}}}
\end{align}
\end{subequations}
$(Z)$ $W$ and $\Gamma$ are respectively the generators of (connected) Greens functions and proper vertex functions.

The background field formalism consists in introducing also a set of classical fields $\beta_i$ and extended functionals (indicated by  \, $\widetilde{}$ \, ):
\begin{align}
\widetilde W\left[ {J;\beta } \right] &= {\mathcal{N}^{ - 1}}\int {\left[ {d\phi } \right]\exp \left\{ {i\int {{d^4}x\left[ {\lag\left( {\phi  + \beta } \right) + {J_i}{\phi _i}} \right]} } \right\}} \label{eq1}
\end{align}
\begin{subequations}
\begin{align}
\widetilde Z\left[ {J;\beta } \right] &=  - i\ln \widetilde W\left[ {J;\beta } \right] = \widetilde \Gamma \left[ {\widetilde \Phi ;\beta } \right] + \int {{d^4}x{J_i}{{\widetilde \Phi }_i}} \\
{\widetilde \Phi _i} &= \frac{{\delta \widetilde Z}}{{\delta {J_i}}}, \qquad {J_i} =  - \frac{{\delta \widetilde \Gamma }}{{\delta {{\widetilde \Phi }_i}}}
\end{align}
\label{eqn4}
\end{subequations}

Here $(\widetilde{Z})$ $\widetilde{W}$ and $\widetilde{\Gamma}$ are the generators of (connected) Greens functions and proper vertex functions respectively in the presence of the background field, the moments being with respect to $J$ and $\widetilde{\Phi}$. Why does one want to do this? Clearly ${\left. {\widetilde \Gamma \left[ {\widetilde \Phi ;\beta } \right]} \right|_{\beta  = 0}} = \Gamma \left[ \Phi  \right]$ but also by translational invariance of the measure $\phi \to \phi - \beta$ in \eqref{eq1}
\begin{align}
\widetilde Z\left[ {J;\beta } \right] = Z\left[ J \right] - \int {{d^4}x{J_i}{\beta _i}} 
\end{align}
which implies $\widetilde \Phi  = \Phi  - \beta $ and therefore
\begin{align}
\widetilde \Gamma \left[ {\widetilde \Phi ;\beta } \right] = Z\left[ J \right] - \int {{d^4}x{J_i}\left( {{{\widetilde \Phi }_i} + {\beta _i}} \right) = } \Gamma \left[ {\widetilde \Phi  + \beta } \right]
\end{align}
so 
\begin{align}
\Gamma \left[ \beta  \right] = \widetilde \Gamma \left[ {0;\beta } \right]:
\end{align}
the usual effective action is calculable from vacuum diagrams (no external legs) in the presence of the background field. 

In practice of course except for very simple background fields -- such as $\beta=$ const. which gives the effective potential $V(\beta)$ -- one cannot carry out the exact evaluation of $\widetilde \Gamma \left[ {0;\beta } \right]$, but usually one only needs the moments of $\Gamma \left[ \beta  \right]$ e.g. to construct the S-matrix. Then in \eqref{eq1} one may expand the action in the exponential in powers of $\beta$
\begin{align}
S\left[ {\phi  + \beta } \right] = \int {{d^4}xL\left( {\phi  + \beta } \right)}  = S\left[ \phi  \right] + \int {{d^4}x{\beta _i}{{\left. \left({\frac{{\delta S}}{{\delta {\beta _i}}}}\right) \right|}_{{\beta _i} = 0}}}  +  \ldots 
\label{eq2}
\end{align}
yielding a set of Feynman rules in which the (classical background) quantum field $(\beta)$ $\phi$ only appears on (external) internal lines - note that for the proper vertex functions this means that per definition one may discard terms linear in $\phi$ in \eqref{eq2}

Nor is this all; suppose $\lag(\phi)$ is invariant under the infinitesimal transformations
\begin{align}
\delta {\phi _i} =  - i{\theta ^a}{\left( {{t_a}} \right)_{ij}}{\phi _j}
\label{eqn29}
\end{align}
of some global group $G$. Then under the simultaneous transformations
\begin{align}
\delta {\beta _i} =  - i{\theta ^a}{\left( {{t_a}} \right)_{ij}}{\phi _j} \quad , \quad 
\delta {J_i} =  - i{\theta ^a}{\left( {{t_a}} \right)_{ij}}{J_j} \label{eqn210}
\end{align}
it is easy to see that $S\left[ {\phi  + \beta } \right]$ and $\int {{d^4}x{J_i}{\phi _i}} $ are invariant so 
\begin{align}
\widetilde W\left[ {J;\beta } \right] = \widetilde W\left[ {J + \delta J;\beta  + \delta \beta } \right]
\end{align}
which implies
\begin{align}
\widetilde \Gamma \left[ {\widetilde \Phi ;\beta } \right] + \int {{d^4}x{J_i}{{\widetilde \Phi }_i}}  = \widetilde \Gamma \left[ {\widetilde \Phi ;\beta  + \delta \beta } \right] + \int {{d^4}x\left( {{J_i} + \delta {J_i}} \right){{\widetilde \Phi }_i}} 
\end{align}
or using \eqref{eqn4} and \eqref{eqn210}
\begin{align}
\widetilde \Gamma \left[ {\widetilde \Phi ;\beta } \right] = \widetilde \Gamma \left[ {\widetilde \Phi  + \delta \widetilde \Phi ;\beta  + \delta \beta } \right]
\end{align}
Hence holding $\widetilde{\Phi}$ to vanish the effective action is invariant:
\begin{align}
\Gamma \left[ \beta  \right] = \Gamma \left[ {\beta  + \delta \beta } \right] \label{eqn5}
\end{align}
In turn thus implies the GWI's as moments of
\begin{align}
0 = \int {{d^4}x\frac{{\delta \Gamma }}{{\delta {\beta _i}}}{{\left( {{t_a}} \right)}_{ij}}{\beta _i}} 
\end{align}

The last may be used to construct a simple proof of renormalizability when $\lag=\lag_H$ above with $G$ global: $\Gamma[\beta]$ may also be expanded locally, 
\begin{align}
\Gamma \left[ \beta  \right] = \int {{d^4}x\left[ { - V\left( {{\phi ^T}\phi } \right) + \frac{1}{2}{Z_{ij}}\left( \beta  \right){\beta _{i,\mu }}\beta _j^{,\mu } +  \ldots } \right]}
\end{align}
the ellipsis standing for higher derivatives. By standard power counting analysis the superficial degree of divergence $D$ is $D=4-E_\beta -k$ where $E_\beta$ is the number of $\beta-$field lines and $k$ the number of derivatives. In view of \eqref{eqn5} it follows immediately that the possible divergences are contained in  
\begin{align}
{\Gamma _D}\left[ \beta  \right] = \int {{d^4}x\left[ {\frac{1}{2}\delta {\mu ^2}{\beta ^T}\beta  - \frac{1}{4}\left( {1 - {Z_\lambda }} \right){{\left( {{\beta ^T}\beta } \right)}^2} + \frac{1}{2}\left( {1 - {Z_\beta }} \right)\beta _{,\mu }^T{\beta ^{,\mu }}} \right]}
\end{align}
which are cancelled by adding counter terms: $\lag \to \lag+\lag_{CT}$
\begin{align}
{L_{CT}} = \frac{1}{2}\left( {{Z_\beta } - 1} \right)\phi _{,\mu }^T{\phi ^{,\mu }} - \frac{1}{2}\delta {\mu ^2}{\phi ^T}\phi  - \frac{1}{4}\lambda \left( {{Z_\lambda } - 1} \right){\left( {{\phi ^T}\phi } \right)^2}
\label{eqn218}
\end{align}
equivalent to the multiplicative renormalizations 
\begin{align}
{\phi _0} = \sqrt {{Z_\beta }} \phi  \quad ,\quad
{\beta _0} = \sqrt {{Z_\beta }} \beta \quad ,\quad
\mu _0^2 = \left( {{\mu ^2} + \delta {\mu ^2}} \right)/{Z_\beta } \quad ,\quad
{\lambda _0} = \lambda {Z_\lambda }/Z_\beta ^2
\end{align}

Note particularly that nothing in the proof outlined depends upon whether $\mu^2 >0$ (Wigner-Weyl phase) or $\mu^2 <0$ (Nambu-Goldstone phase) -- provided only that one chooses an appropriate renormalization scheme (such as minimal subtraction) which is non-singular as $\mu^2$ is continued from positive to negative values the renormalizability of the symmetric phase implies the renormalizability of the asymmetric phase.

\section{The Background Field Gauge}
For global symmetries as in the preceding section the invariance of the effective action and the renormalizability of the (a-)symmetric phase can be proven without resort to the background field. Where the method begins to come into its own is in the Yang-Mills theory where the definition of the generating functionals require that one break the gauge invariance erstwhile the $A$ field propagator is undefined. The magic of the background field method is that one can break the \textbf{quantum} field invariance while retaining background field gauge invariance provided one is clever in choosing the gauge fixing \cite{bdw,tHooft2,labb1,labb2}.

One begins with 
\begin{align}
W\left[ {J;B} \right] = {\mathcal{N}^{ - 1}}\int {\left[ {dA} \right]\exp \left\{ {i\int {{d^4}x\left[ {{\lag_{eff}}\left( {A;B} \right) - J_a^\mu A_\mu ^a} \right]} } \right\}}
\end{align}
where $B_\mu ^a$ is the background field to $A_\mu ^a$ and
\begin{align}
{\lag_{eff}}\left( {A;B} \right) =  - \frac{1}{4}F_{\mu \nu }^a\left( {A + B} \right){F^{a\mu \nu }}\left( {A + B} \right) + {\lag_{GF}} + {\lag_{FPG}}
\end{align}
In the absence of gauge fixing ($\lag_{GF}=\lag_{FPG}=0$) there is invariance under the infinitesimal transformations
\begin{align}
\delta A_\mu ^a = i{\left( {{T_c}} \right)^{ab}}{\theta _b}A_\mu ^c \quad ,\quad
\delta J_a^\mu  = {C_{abc}}{\theta ^b}J_c^\mu \quad ,\quad
\delta B_\mu ^a = \frac{1}{g}D_\mu ^{ab}\left( B \right){\theta _b}
\label{eqn6}
\end{align}
which would be broken by the usual choice $f^a(A)=\partial^\mu A_\mu ^a$. On the other hand, replacing the ordinary derivative by the gauge covariant derivative with respect to the background field
\begin{align}
{G^a}\left( {A;B} \right) = D_\mu ^{ab}\left( B \right){A^{b\mu }}
\label{eqn34}
\end{align}
one observes that $G^a$ transforms as the adjoint representation of the gauge group
\[G \to (I-i\theta^a T_a)G\]
so 
\begin{align}
\lag_{GF} = \lag_{BGF} = -\frac{1}{2\xi}(G^a)^\dagger G^a
\end{align}
is invariant under \eqref{eqn6}. The Faddeev-Popov operator transforms similarly 
\begin{align*}
M_G \to (I-i\theta^a T_a)M_G (I+i\theta^a T_a)
\end{align*}

hence $|M_G|$ is invariant under \eqref{eqn6} as is 
\begin{align}
\lag_{FPG} = \lag_{BFPG} = C_a ^\dagger D_\mu ^{ab}(B) D^{bd\mu}(A+B)C_d
\end{align}
Then by identical steps as in Section 2,
\begin{align}
\widetilde \Gamma \left[ {0;B} \right] = \Gamma \left[ B \right] = \Gamma \left[ {B + \delta B} \right] \label{eqn7}
\end{align}
yielding the GWI's as moments of
\begin{align}
0 = \int {{d^4}xD_\mu ^{ab}\left( B \right)\frac{{\delta \Gamma }}{{\delta B_\mu ^b}}} 
\end{align}

Due to the background field gauge invariance \eqref{eqn7} and power counting, $D=4-E_B - k$, one immediately obtains that the divergences are cancelled by the gauge invariant counter-terms
\begin{align}
{\lag_{CT}}\left( A \right) =  - \frac{1}{4}\left( {{Z_b} - 1} \right)F_{\mu \nu }^a\left( A \right){F^{a\mu \nu }}\left( A \right)
\end{align}
Moreover, defining the bare field and coupling 
\begin{align}
B_0^{a\mu } = \sqrt {{Z_B}} {B^{a\mu }} \quad , \quad
{g_0} = {Z_g}g
\end{align}
in order that $\lag_{0YM}(B_0)$ be invariant one has in view of the nonlinear terms the identity 
\begin{align}
Z_g \sqrt{Z_B}=1
\end{align}
In turn this means that to calculate the $\beta$ function  for QCD in the background field formalism one only needs the two point function for the $B$ field, as opposed to the standard methods where one requires two and three point proper vertices. Beyond one loop the resulting computational advantages are enormous \cite{labb1,labb2}

\section{THE BFG for spontaneously broken theories}
With the foregoing in mind we now return to the problem of spontaneously broken Yang-Mills theories. Let us introduce background fields $\beta$ and $B$ for $\phi$ and $A$, and consider 
\begin{align}
\widetilde W\left[ {J;B,\beta } \right] = {N^{ - 1}}\int {\left[ {dA} \right]\left[ {d\phi } \right]\exp \left\{ {i\int {{d^4}x\left[ {{\lag_{eff}}\left( {A,\phi ;B,\beta } \right) + {J_i}{\phi _i} - J_a^\mu A_\mu ^a} \right]} } \right\}} 
\label{eqn41}
\end{align}
where 
\begin{align}
{\lag_{eff}}\left( {A,\phi ;B,\beta } \right) = \lag\left( {A + B,\phi  + \beta } \right) + {\lag_{GF}} + {\lag_{FPG}}
\end{align}
and $\lag$ is as seen in \eqref{eqn8} to \eqref{eqn9}. With $\lag_{GF}=\lag_{FPG}=0$ \eqref{eqn41} is invariant under the simultaneous infinitesimal local transformations \eqref{eqn29}, \eqref{eqn210} and \eqref{eqn6}, and we seek to choose the (quantum) gauge fixing so as to preserve this feature. Now by it self this would be accomplished by \eqref{eqn34} just as in the pure gauge theory.

There is however a problem: $\lag\left( {A + B,\phi  + \beta } \right)$ contains a piece
\begin{align*}
\frac{1}{2}igA_\mu ^a{\partial ^\mu }\left( {{\phi ^T}{t_a}\beta  - {\beta ^T}{t_a}\phi } \right)
\end{align*}

The proper vertices are found by expanding 
\begin{align}
\Gamma \left[ {B,\beta } \right] = \Gamma \left[ {B + \delta B,\beta  + \delta \beta } \right]
\label{eqn43}
\end{align}
about $B^a _\mu=0, \beta=v$ with $v$ a constant found by minimizing the effective potential
\begin{align}
\Gamma \left[ {0,v} \right] =  - \int {{d^4}xV\left( {{v^T}v} \right)} 
\end{align}
Consequently in the broken symmetry phase there occurs a mixing between the quantum gauge and would-be Goldstone fields.

Taking a clue \cite{nt2} from 't Hooft, the way out of this difficulty is apparent: we introduce a modified background field gauge fixing function, replacing $G^a$ of \eqref{eqn34} by
\begin{align}
{H^a}\left( {A,\phi ;B,\beta } \right) = {D^{ab\mu }}\left( B \right)A_\mu ^b - i\xi {\phi ^T}{t_a}\beta 
\label{eqn45}
\end{align}
which also transforms as the adjoint representation of the gauge group under \eqref{eqn29} and \eqref{eqn6}, insuring the background field invariance of 
\begin{align}
{\lag_{GF}} = {\lag_{BGF}} =  - \frac{1}{{2\xi }}{\left( {{H^a}} \right)^\dag }{H^a}
\label{eqn46}
\end{align}
as well as 
\begin{align}
{\lag_{FPG}} = {\lag_{BFPG}} = C_a^\dag D_\mu ^{ab}\left( B \right){D^{bd\mu }}\left( {A + B} \right){C_d} + \xi {g^2}{\phi ^T}{t_d}{t_a}\beta C_a^\dag {C_d}
\end{align}
Crucially \eqref{eqn45} and \eqref{eqn46} also assure the cancellation of the quantum mixing piece in the broken symmetry phase, $v\ne 0$.

Let us see what we have accomplished; first, the internal ($A$ field) gauge propagator assumes the form \eqref{eqn115} and the GWI's are from \eqref{eqn43}
\begin{align}
\int {{d^4}xD_\mu ^{ab}\left( B \right)\frac{{\delta \Gamma }}{{\delta B_\mu ^b}}}  = ig\int {{d^4}x\frac{{\delta \Gamma }}{{\delta {\beta _i}}}{{\left( {{t_a}} \right)}_{ij}}{\beta _j}} 
\end{align}
The background field invariance together with power counting $D=4-E_B - E_\beta - k$ means that the proof of renormalizability goes through for the coupled Higgs scalar-gauge field system as for its component parts (replacing the ordinary by the gauge covariant derivative in \eqref{eqn218}). Thus \textbf{renormalizability is manifest}.

Second, because we have retained the background field gauge invariance \eqref{eqn43}, we can perform a unitary gauge transformation on the background fields:
\begin{subequations}
\begin{align}
\beta  \to \exp \left( { - i{\theta ^a}{t_a}/v} \right)\left( {v + \eta } \right) = U\left( \theta  \right)\left( {v + \eta } \right)\\
{t_a}B_\mu ^a \to U\left( \theta  \right){t_a}B_\mu ^a{U^\dag }\left( \theta  \right) - \frac{i}{g}U\left( \theta  \right){\partial _\mu }{U^\dag }\left( \theta  \right)
\end{align}
\end{subequations}
so 
\begin{align}
\Gamma \left[ {B,\beta } \right] \to \Gamma \left[ {B,v + \eta } \right]
\end{align}
Using $\Gamma \left[ {B,v + \eta } \right]$ to generate the proper vertex functions the background would be Goldstone bosons never appear in the S-matrix, the massive background gauge field propagator being that of \eqref{eqn111} plus loop corrections. hence also \textbf{unitarity is manifest}

\section{Conclusions}
In this paper I have shown how the background field formalism and a simple modification of the conventional background field gauge allows one to rectify the usual schism between unitarity and renormalizability for spontaneously broken gauge theories. It has further been demonstrated that the background field method leads to much simpler and elegant proofs than in the conventional approach. 

It is also to be hoped that, esoterics aside, the method leads to computational simplifications in spontaneously broken theories in the same that it has for e.g. the $\beta$ function in unbroken Yang-Mills theory. This aspect is currently under investigation

\section*{Acknowledgements}
I would like to thank J. Reid for having first introduced me to the elegance of the background field method.


\begin{thebibliography}{9}
\bibitem{c3} 
	This includes a part $-i(r/2) \delta^{(n)}(0)\ln(1+\eta/v)^2$ arising from the transformation of the measure where $r$ is the order of the coset $G/G'$. In $n=4$ dimensions this piece effects the cancellation of quartic divergences, $\delta^{(n)}(0)=\int d^4 k/(2\pi)^4$, while it vanishes in dimensional regularization.
\end{thebibliography}
\end{document}